\documentclass{article}

\usepackage{arxiv}
\usepackage[utf8]{inputenc}
\usepackage[T1]{fontenc}
\usepackage{hyperref}
\hypersetup{hypertexnames=false}
\usepackage{url}
\usepackage{graphicx}%
\usepackage{multirow}%
\usepackage{amsmath,amssymb,amsfonts}%
\usepackage{amsthm}%
\usepackage{mathrsfs}%
\usepackage{xcolor}%
\usepackage{textcomp}%
\usepackage{booktabs}%
\usepackage{algorithm}%
\usepackage{algorithmicx}%
\usepackage{algpseudocode}%
\usepackage{listings}%
\usepackage{array}%
\usepackage{makecell}%
\usepackage{authblk}
\usepackage{cite}

\graphicspath{{figures/}}
\newcommand{\botrule}{\bottomrule}
\let\standardabstract\abstract
\let\endstandardabstract\endabstract
\renewcommand{\abstract}[1]{\standardabstract #1\endstandardabstract}
\renewcommand{\keywords}[1]{\par\medskip\noindent\textbf{Keywords:} #1}

\begin{document}

\title{Multi-Catheter Digitization in Brachytherapy via Few-Shot Synthetic-to-Real Learning and Structure-Aware Tracking}

\author[1]{Zhuo Xiao}
\author[1,2]{Bo Liu\thanks{Corresponding author: \texttt{bo.liu@buaa.edu.cn}}}
\author[1]{Jingjing Wang}
\author[1]{Qinglong Yao}
\author[3]{Haitao Sun}
\author[1]{Fugen Zhou}
\author[3]{Junjie Wang}
\author[4]{Qiuwen Wu}
\author[3]{Ping Jiang}

\affil[1]{Image Processing Center, Beihang University, Beijing 100191, People's Republic of China}
\affil[2]{State Key Laboratory of High-Efficiency Reusable Aerospace Transportation Technology, Beijing 102206, People's Republic of China}
\affil[3]{Department of Radiation Oncology, Peking University Third Hospital, Beijing 100191, People's Republic of China}
\affil[4]{Department of Radiation Oncology, Duke University Medical Center, Durham, North Carolina 27708, USA}
\date{}
\maketitle

\abstract{\textbf{Purpose:} Accurate catheter digitization in CT-guided interstitial brachytherapy (ISBT) is a critical yet time-consuming bottleneck, particularly for complex implant configurations. This study aimed to develop and validate a data-efficient, physics-guided framework for automated multi-catheter digitization requiring minimal clinical annotation. \textbf{Methods:} We propose a two-stage pipeline that addresses both data scarcity and the geometric complexities associated with catheter adhesion. First, an implant region-aware network (IRNet) was pretrained on synthetic CT volumes simulating localized metallic signatures, followed by highly efficient 10-shot fine-tuning on clinical data. Second, a structure-aware reconstruction module employed a direction-constrained 3D Hough transform and synchronous, physics-constrained inward tracking to disentangle adherent catheter trajectories. The framework was evaluated using patient-level five-fold cross-validation on 203 clinical treatment fractions from 38 patients. \textbf{Results:} The 10-shot fine-tuned IRNet demonstrated high geometric reliability, achieving a 95th-percentile Hausdorff distance (HD95) of 0.853 $\pm$ 0.362 mm. In fully automated end-to-end evaluation, the framework achieved an F1 score of 0.891 $\pm$ 0.178, with submillimeter shaft and tip digitization errors of 0.334 $\pm$ 0.367 mm and 0.896 $\pm$ 0.680 mm, respectively. Notably, the method maintained robust tracking performance, with an F1 score of 0.843 $\pm$ 0.190, in the challenging subgroup characterized by severe catheter adhesion. The complete automated workflow required approximately 11.6 s per case. \textbf{Conclusion:} By synergizing data-efficient deep learning with physics-guided structural tracking, the proposed framework provides a robust and scalable solution for multi-catheter digitization. It substantially reduces the manual annotation burden and effectively resolves complex catheter adhesions, demonstrating strong potential for integration into time-sensitive clinical ISBT workflows.}


\section{Introduction}\label{sec:introduction}

Cervical cancer remains the fourth leading cause of cancer-related mortality among women globally\cite{brayGlobalCancerStatistics2024}. As a critical intervention for advanced or recurrent gynecological malignancies, high-dose-rate (HDR) interstitial brachytherapy (ISBT) has demonstrated superior local control rates\cite{cibulaESGOESTROESP2023,potter2021mri}. In contemporary image-guided brachytherapy, interstitial catheters are particularly important for large, irregular, parametrial, or vaginally extended tumors, where conventional intracavitary applicators alone often fail to provide adequate target coverage without exceeding tolerance doses to adjacent organs at risk\cite{GPM10968}. Consequently, the clinical benefit of combining intracavitary and interstitial techniques has been increasingly recognized and adopted in modern cervical cancer treatment paradigms\cite{yangBrachytherapyCervicalCancer2024}.

In CT-guided HDR-ISBT, the accuracy of treatment planning fundamentally relies on the precise identification and reconstruction of implanted catheters. These reconstructed paths define the source trajectory and valid dwell locations, thereby directly dictating subsequent dose calculation and optimization\cite{mouryaEvolutionBrachytherapyApplicators2021,beriwal2012american}. Current AAPM practice guidelines identify applicator, needle, and catheter reconstruction as a technically challenging and safety-critical component of HDR brachytherapy planning\cite{Richardson2025MPPG13B}. Because HDR brachytherapy is characterized by steep dose gradients, even small geometric errors in catheter reconstruction can translate into clinically critical dosimetric deviations. Previous studies have shown that both random and systematic reconstruction uncertainties substantially compromise dose-volume histogram (DVH) parameters, underscoring the absolute necessity for highly reliable catheter digitization\cite{tanderupConsequencesRandomSystematic2008}.

However, the accurate digitization of multiple interstitial catheters on post-implant CT remains a significant clinical bottleneck. During perineal implantation, metallic catheters are densely distributed within a highly restricted anatomical region. They often exhibit adhesion trajectories and generate severe metal-induced streak artifacts that obscure surrounding tissues. Under these adverse imaging conditions, manual digitization is highly operator-dependent, labor-intensive, and prone to intra- and inter-observer variability. For complex implant configurations, this manual process can consume tens of minutes of valuable clinical time\cite{wangAttentionGatedDeepLearning2024,shaaerDeepLearningAssistedAlgorithm2022}, becoming increasingly unsustainable within time-sensitive workflows. While hardware-based tracking technologies, particularly electromagnetic tracking, offer an alternative \cite{poulinFastAutomaticAccurate2015,thoTechnicalNoteEM2018,sauerElectromagneticTrackingInterstitial2022,cantin2024proof,DEUFEL2024676}, their broader clinical adoption remains limited by the need for dedicated hardware, registration with imaging systems, susceptibility to electromagnetic and metallic interference, and incomplete integration into routine brachytherapy workflows \cite{Beaulieu2025TG317}. Consequently, image-based automatic digitization represents the most practical and scalable direction for routine ISBT workflows.

Most existing image-based automatic digitization methods follow a two-stage paradigm: catheter segmentation followed by trajectory reconstruction\cite{moradiFullyAutomaticReconstruction2025,quetin2025automatic,jungDeeplearningAssistedAutomatic2019}. While deep learning models\cite{moradiFullyAutomaticReconstruction2025,quetin2025automatic,jungDeeplearningAssistedAutomatic2019,zhangAutomaticSegmentationApplicator2020,zhengAutomaticNeedleDetection2021,mehrtashAutomaticNeedleSegmentation2019,zhouDeepLearningbasedAutomatic2024,aleongRapidMulticatheterSegmentation2024,daiAutomaticMulticatheterDetection2020,wangDeepLearningApplications2020,zhangMultineedleLocalizationAttention2020,zhangAutomaticMultineedleLocalization2020} have substantially improved segmentation robustness over conventional image-processing techniques\cite{diseDevelopmentEvaluationAutomatic2015,hrinivichSimultaneousAutomaticSegmentation2017} across various modalities, this paradigm still faces two critical bottlenecks that limit its clinical translation. First, the performance of deep learning-based segmentation is inherently bounded by the availability of large, high-quality annotated datasets\cite{quetin2025automatic,11523703}. In dense multi-catheter ISBT, acquiring comprehensive manual annotations is not only tedious but also prone to human error due to severe artifacts. Existing data augmentation strategies, such as copy-paste methods\cite{ghiasiSimpleCopyPasteStrong2021} or standard generative models\cite{11367016}, often fail to capture the complex, physically grounded CT appearance of metallic implants and their intricate artifacts. This data-scarcity problem poses a significant barrier to the adoption of deep learning models, particularly for institutions with limited clinical case volumes that struggle to amass sufficiently large training cohorts. Second, trajectory reconstruction in dense and adhesion multi-catheter settings remains an unresolved algorithmic challenge. Although simple and well-separated cases can often be handled by connected-component analysis\cite{zhangMultineedleLocalizationAttention2020} or slice-wise tracking\cite{moradiFullyAutomaticReconstruction2025,zhangAutomaticMultineedleLocalization2020}, these strategies tend to fail when adjacent catheters touch, overlap, or cross, because voxel assignment becomes ambiguous. To our knowledge, the methods proposed by Jung et al.\cite{jungDeeplearningAssistedAutomatic2019} and Quetin et al.\cite{quetin2025automatic} are among the few published catheter digitization approaches that explicitly incorporate dedicated mechanisms for handling catheter adhesion. Jung et al. formulated the separation of adherent catheters as a dedicated optimization problem, whereas Quetin et al. introduced iterative separation criteria to disentangle adherent trajectories. These methods therefore provide the most directly relevant baselines for evaluating reconstruction robustness in catheter-adhesion cases. Nevertheless, these existing approaches often rely heavily on local continuity assumptions. When multiple catheters merge across continuous CT slices and local visual features are heavily corrupted by artifacts, these localized tracking algorithms struggle with instance separation, leading to trajectory deviations or premature termination.

To address these challenges, we propose a data-efficient and structure-aware multi-catheter digitization framework that seamlessly couples few-shot semantic segmentation with physics-guided trajectory reconstruction. Our methodology balances data economy with geometric robustness to resolve the distinct bottlenecks in contemporary brachytherapy workflows. The main contributions of this work are summarized as follows.

First, we formulate data-efficient catheter segmentation as a physics-guided few-shot domain adaptation (P-FSDA) problem. Physically synthetic catheter configurations and metal artifacts are used to provide source-domain supervision, followed by limited adaptation to clinical CT images. We further develop an Implant Region-aware Network (IRNet), which enforces spatial compactness via an auxiliary gating module to suppress anatomically implausible false positives. Systematic evaluations across progressive few-shot regimes demonstrate that a minimal budget of 10 clinical fractions effectively bridges the synthetic-to-real gap, markedly improving geometric reliability as reflected by HD95.

Second, we develop a structure-aware trajectory reconstruction algorithm driven by physical insertion priors. By exploiting the undeformed rigid characteristics of external catheter segments via a direction-constrained 3D Hough transform, the pipeline establishes robust initial orientations. These global directions guide a synchronous, physics-constrained inward-tracking process to successfully untangle intersecting centerlines in highly adherent regions.

Third, we validate the end-to-end framework on a comprehensive clinical cohort. The balanced integration of segmentation and tracking provides a robust and computationally efficient approach requiring approximately 11.6 seconds per case, demonstrating strong clinical viability for routine brachytherapy workflows.

\section{Materials and Methods}\label{sec:materials-and-methods}

\begin{figure}[htbp]
\centering
\includegraphics[width=0.95\textwidth]{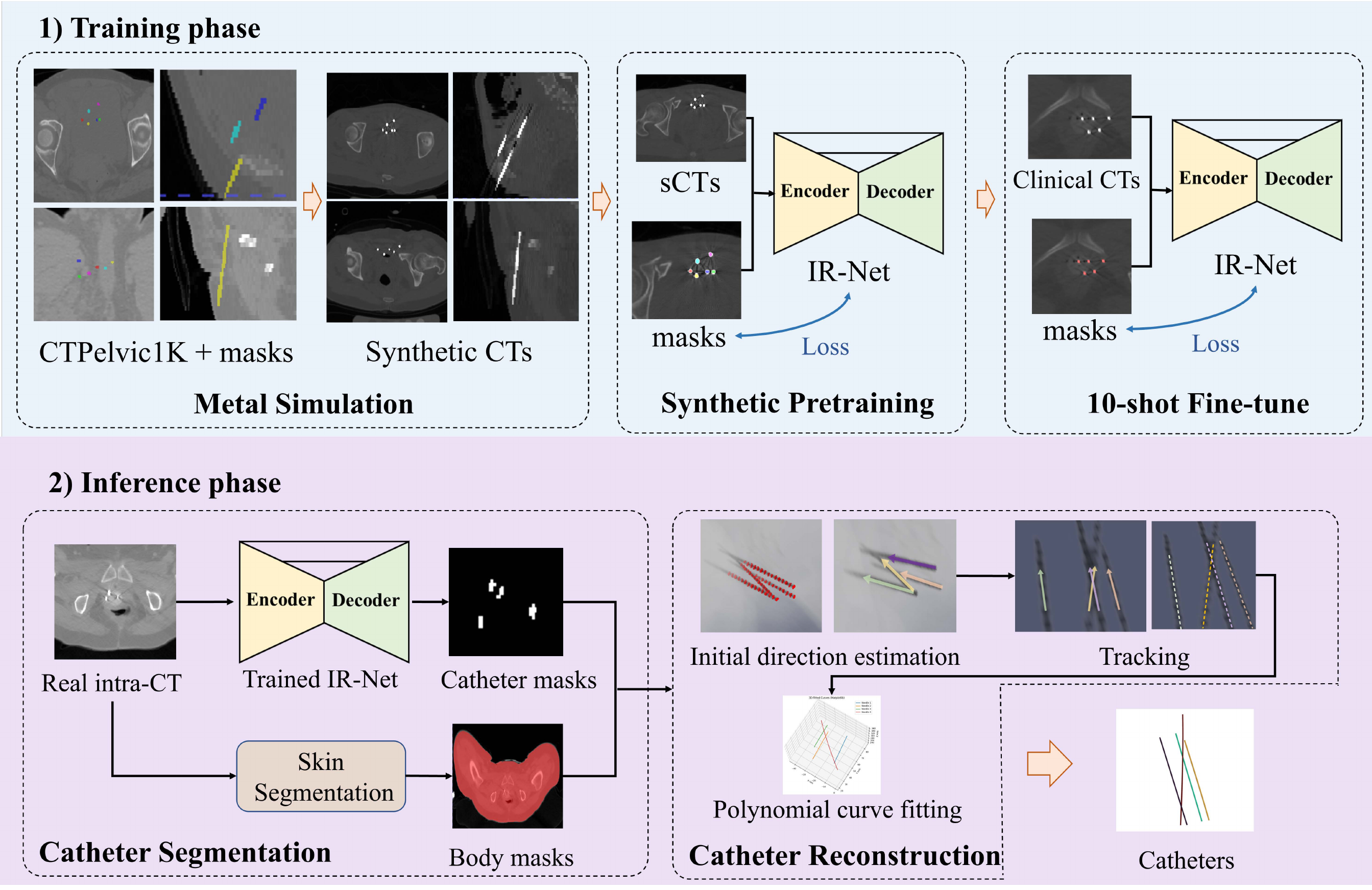}
\caption{Schematic overview of the proposed data-efficient and structure-aware multi-catheter digitization framework, illustrating: (1) physics-driven source-domain pretraining with few-shot clinical adaptation, and (2) target clinical inference utilizing region-gated segmentation and geometry-constrained trajectory reconstruction.}\label{fig:framework}
\end{figure}

The overall architecture of the proposed P-FSDA framework is illustrated in Fig.~\ref{fig:framework}. The multi-catheter digitization workflow is conceptualized into two primary complementary phases: physics-driven source domain pretraining with progressive few-shot adaptation, and automated target clinical inference.

In the first phase, a template-propagated data generation pipeline coupled with slice-wise metal artifact simulation is established to construct a large-scale synthetic dataset with automatically generated voxel-level labels using catheter-free pelvic CTs. IRNet is initially optimized on this source domain to master local metallic and artifact signatures, and is subsequently adapted to the target clinical domain using nested, progressive few-shot subsets ranging from 1 to 20 shots to achieve cross-domain generalization under a highly restricted clinical annotation budget.

In the second phase, the adapted IRNet extracts semantic catheter masks from real clinical post-implant CT scans. These masks, alongside an intensity-thresholded body contour mask, are subsequently channeled into a geometry-guided, structure-aware reconstruction pipeline. This backend pipeline sequentially executes external initial direction estimation via a direction-constrained 3D Hough transform, synchronous physics-constrained tracking, and third-order polynomial regularization to recover distinct instance-level trajectories with high physical fidelity.

The text below details the clinical and simulated data distributions, elaborates on the data-efficient segmentation network, describes the structure-guided reconstruction module, and provides the full implementation parameters.

\subsection{Patient data and preprocessing}\label{sec:patient-data-and-preprocessing}

This retrospective study evaluated 203 fraction CT images from 38 patients undergoing CT-guided interstitial brachytherapy for locally recurrent cervical cancer at Peking University Third Hospital between November 2017 and April 2023. Post-implant CT volumes were acquired using a Brilliance Big Bore scanner (Philips Healthcare; 120 kV, 150 mA). Original in-plane resolutions ranged from 0.80 to 1.29 mm with slice thicknesses of 2 to 5 mm, with a majority of 5 mm. All volumes were resampled to a uniform 5 mm axial resolution for consistent downstream processing.

Implant configurations comprised 4 to 13 catheters per fraction, with a mean of 7.31 catheters. Catheter adhesion was present in 83 of the 203 fractions, highlighting the geometric complexity of the clinical cohort. Ground truth trajectories were delineated by a medical physicist and independently verified and refined by a second. These validated paths were exported as unique instance-level reference labels and subsequently merged into a unified foreground class to provide semantic masks for segmentation evaluation.

To prevent patient-level data leakage, all experiments utilized patient-level five-fold cross-validation, ensuring every fraction was evaluated as unseen test data exactly once. Within each fold, training patients were utilized either for fully supervised baseline training from random initialization or for the few-shot fine-tuning of the synthetic-pretrained IRNet. For the catheter adhesion subgroup analysis, each fraction was included only in the held-out test fold containing its corresponding patient.

For synthetic pretraining, 41 catheter-free pelvic CT volumes from the CTPelvic1K\cite{liuDeepLearningSegment2021} cervix subset, with original in-plane resolution of 0.64--1.13 mm and slice thickness of 0.8 mm, were similarly resampled to a 5 mm axial thickness prior to synthetic mask generation and metal artifact simulation.

\subsection{Physics-guided Few-Shot Domain Adaptation}\label{sec:p-fsda}

Formally, we define the source domain as
\begin{equation}
\mathcal{D}_{s}=\{(X_i^{s},Y_i^{s})\}_{i=1}^{N_s},\quad N_s=123,
\end{equation}
where $X_i^{s}$ represents the synthesized pelvic CT volumes and $Y_i^{s}$ represents the corresponding voxel-level catheter annotations. The target clinical domain is defined as
\begin{equation}
\mathcal{D}_{t}=\{X_j^{t}\}_{j=1}^{N_t}.
\end{equation}
The objective of our P-FSDA framework is to optimize the parameterized mapping $f_{\theta}$ from $\mathcal{D}_{s}$ to $\mathcal{D}_{t}$ utilizing a highly restricted target training subset $\mathcal{D}_{t,\mathrm{train}}\subset\mathcal{D}_{t}$, where $|\mathcal{D}_{t,\mathrm{train}}|=K$. The objective is to learn a segmentation model $f_{\theta}$ that exploits large-scale synthetic supervision and a limited number of annotated clinical fractions to generalize to unseen clinical CT volumes.

Operationally, the proposed framework comprises three sequential stages: construction of a physically realistic synthetic source domain, source-domain pretraining of the implant region-aware network, and full-network adaptation using the $K$-shot clinical subset. Accordingly, the following subsections first describe the generation of paired synthetic CT images and catheter masks, and then introduce the IRNet architecture and its training objective. Detailed pretraining and few-shot adaptation protocols are provided in Section~\ref{sec:training-and-inference}.

\subsubsection{Synthetic CT generation}\label{sec:synthetic-ct-generation}

Training robust deep learning models for catheter segmentation relies on large and well-annotated datasets of post-implant CT images. However, acquiring large real clinical cohorts is difficult, and manually delineating thin interstitial catheters under severe metal artifacts remains time-consuming and prone to inter-observer variability. To reduce reliance on real patient data and manual annotation, we developed a synthetic pretraining pipeline to provide large-scale voxel-level supervision before fine-tuning. This strategy constructs catheter masks on a reference metal-free pelvic CT and subsequently transfers them to new cases via image registration and geometric augmentation, enabling the efficient generation of diverse implant configurations across varying anatomical backgrounds.

For template construction, one 3D CT volume from the cervix subset of CTPelvic1K\cite{liuDeepLearningSegment2021} was selected as the reference case. On this reference CT, 13 virtual catheters were placed at corresponding anatomical locations to form an initial implant configuration. Each catheter was then converted into an individual 3D instance mask, and the resulting set of masks was used as the initial implant template.

To propagate this template to novel anatomical backgrounds, each target catheter-free pelvic CT was rigidly registered to the reference CT. The resulting inverse transformation was then applied to map the catheter masks from the reference space into the target image space, generating an initial catheter layout tailored to each specific target CT. To further expand geometric diversity, these propagated masks underwent targeted data augmentation. Specifically, a random subset of catheters was removed to yield a final count of 4 to 13, mirroring the distribution observed in our clinical cohort. Additionally, minor spatial perturbations were introduced by shifting catheter centroids by up to 2 mm and altering their orientations by up to 5$^\circ$. This synergistic registration and augmentation strategy effectively transformed a single reference template into a multitude of anatomically plausible layouts, capturing a wide variance in catheter counts, spatial distributions, and insertion trajectories.

Finally, realistic post-implant CT images were generated from these augmented layouts utilizing an established metal artifact simulation method\cite{zhangConvolutionalNeuralNetwork2018}. By digitally embedding the simulated metallic catheters into their corresponding catheter-free CT volumes, this procedure yielded paired synthetic CT scans alongside precise, voxel-level annotations, thereby providing a comprehensive synthetic supervised dataset for pretraining the segmentation network prior to few-shot clinical adaptation.

\subsubsection{Implant region-aware segmentation network}\label{sec:implant-region-aware-segmentation-network}

In post-implant CT images, non-target high-density or metallic structures may exhibit image intensities similar to those of interstitial catheters, causing conventional segmentation networks to produce spurious catheter predictions outside the true implant region. In clinical practice, however, implanted catheters are usually spatially clustered within a relatively compact treatment region. Motivated by this structural prior, we designed a 2D implant region-aware network, termed IRNet, to explicitly estimate the probability of the implant region. The estimated implant-region heatmap serves as an auxiliary spatial prior that guides catheter segmentation by enhancing decoder features in anatomically plausible implant areas while suppressing irrelevant activations in distant background regions.

\begin{figure}[htbp]
\centering
\includegraphics[width=0.95\textwidth]{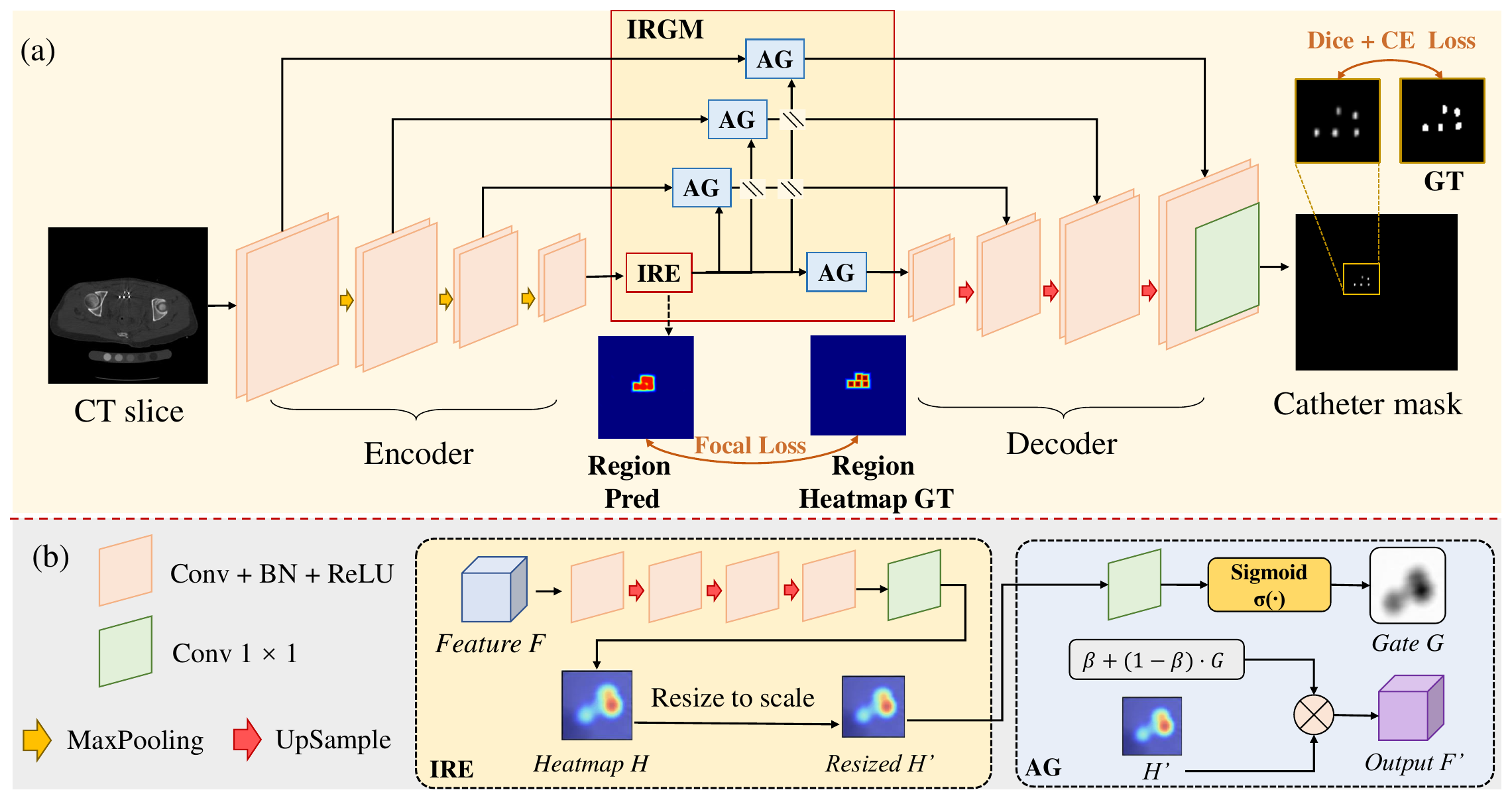}
\caption{Architecture of the proposed Implant Region-aware Network (IRNet). (a) Overall architecture. The implant region gating module (IRGM) predicts a single implant-region heatmap from the deepest encoder feature and injects this shared region prior into multiple decoder stages through scale-adaptive gating. (b) Structures of the basic convolution block and the IRGM.}\label{fig:irnet}
\end{figure}

\noindent\textbf{Network architecture.} As illustrated in Fig.~\ref{fig:irnet}, IRNet adopted a lightweight U-Net-like encoder-decoder architecture. The network used a single-channel CT slice as input and produced a two-channel semantic segmentation output corresponding to background and catheter foreground. The basic convolutional block consisted of two consecutive 3 $\times$ 3 convolutional layers, each followed by batch normalization and ReLU activation. The encoder contained four downsampling stages with a base channel number of 32, while the decoder progressively restored the spatial resolution using transposed convolutions and skip connections.

To provide an explicit implant-region prior, an implant region gating module (IRGM) was integrated into the network. As shown in Fig.~\ref{fig:irnet}, the IRGM consists of two main functional components: an implant region estimator (IRE) and multiple adaptive gating (AG) blocks. The IRE is attached to the deepest encoder stage to extract high-level contextual information and predicts a single-channel heatmap representing the clinically plausible implant region. This heatmap is then distributed to different decoder stages, where it is resized according to the spatial resolution of each decoder feature map. Consequently, the same global implant-region prior is injected into the decoder in a scale-adaptive manner.

The AG blocks are designed to adaptively modulate decoder features according to both the predicted implant-region prior and the feature scale at which the modulation is performed. Let $F_s$ denote the decoder feature at scale $s$ and $H$ denote the original implant-region heatmap predicted by IRE. For each AG block, $H$ is first resized to the spatial resolution of $F_s$, yielding a scale-specific heatmap $H_s$. The resized heatmap is then transformed into a gating map using a 1 $\times$ 1 convolution followed by a sigmoid activation:

\begin{equation}
G_s = \sigma\left(\operatorname{Conv}_{1\times1}(H_s)\right).
\end{equation}

The decoder feature is subsequently modulated by a residual gating operation:

\begin{equation}
F'_s = F_s \odot \left[\beta + (1-\beta)G_s\right],
\end{equation}

where $\odot$ denotes element-wise multiplication and $\beta$ is a residual coefficient introduced to avoid excessive suppression of low-response regions. In this study, $\beta$ was set to 0.1. Through this formulation, the implant-region prior guides the decoder to focus on anatomically valid areas while suppressing false-positive activations in the background.

\noindent\textbf{Loss function.} The reference heatmap was generated from the binary catheter mask to provide soft supervision for implant-region estimation. Let $M$ denote the binary catheter mask, where foreground pixels correspond to catheter regions. For each pixel $x$, the shortest Euclidean distance to the catheter foreground was defined as

\begin{equation}
d(x,M)=\min_{y\in M}\|x-y\|_2 .
\end{equation}

The reference heatmap $H(x)$ was constructed by assigning a value of 1 to the catheter foreground and applying a Gaussian decay to the surrounding pixels:

\begin{equation}
H(x)=\begin{cases}
1, & x\in M,\\
\exp\left(-\frac{d(x,M)^2}{2\sigma^2}\right), & x\notin M\ \mathrm{and}\ d(x,M)\le R,\\
0, & \mathrm{otherwise}.
\end{cases}
\end{equation}

where $R$ is the maximum support radius, and $\sigma$ controls the spatial decay rate. This formulation assigns the highest confidence to the annotated catheter pixels and provides gradually decreasing soft labels in their surrounding neighborhood. In our experiments, the hyperparameters $R$ and $\sigma$ were empirically set to 10 and 4.0 mm, respectively.

The total training objective was formulated as a weighted sum of two components: a segmentation loss $L_{\mathrm{seg}}$ and a heatmap supervision loss $L_{\mathrm{hm}}$. The segmentation loss was defined as a compound objective function combining the cross-entropy (CE) loss $L_{\mathrm{CE}}$ and the Dice loss $L_{\mathrm{Dice}}$:

\begin{equation}
L_{\mathrm{seg}}=L_{\mathrm{CE}}+L_{\mathrm{Dice}}.
\end{equation}

To effectively handle continuous target probabilities $H(x)\in[0,1]$, the heatmap supervision loss $L_{\mathrm{hm}}$ was defined as a focal loss\cite{linFocalLossDense2020}. Let $\hat{H}(x)=\sigma(z(x))$ denote the predicted heatmap probability obtained from the heatmap logit $z(x)$. The loss is computed as

\begin{equation}
\begin{aligned}
L_{\mathrm{hm}}=-\frac{1}{|\Omega|}\sum_{x\in\Omega}\bigl[&\alpha H(x)(1-\hat{H}(x))^{\gamma}\log \hat{H}(x)\\
&+(1-\alpha)(1-H(x))\hat{H}(x)^{\gamma}\log(1-\hat{H}(x))\bigr],
\end{aligned}
\end{equation}

where $\Omega$ denotes the set of image pixels, and $\alpha$ and $\gamma$ are focal hyperparameters. In this study, $\alpha=0.25$ and $\gamma=2.0$.

The overall loss function was written as

\begin{equation}
L=\lambda_{\mathrm{seg}}L_{\mathrm{seg}}+\lambda_{\mathrm{hm}}L_{\mathrm{hm}}.
\end{equation}

In this study, the weighting coefficients were empirically set to $\lambda_{\mathrm{seg}}=1$ and $\lambda_{\mathrm{hm}}=0.5$.

\subsection{Target Clinical Inference and Structure-Aware Reconstruction}\label{sec:target-clinical-inference}

\subsubsection{Catheter segmentation and body contour extraction}\label{sec:catheter-segmentation-and-body-contour-extraction}

During inference, the trained IRNet was applied directly to clinical post-implant CTs to extract semantic catheter masks. Concurrently, a body contour mask was generated to delineate intra- and extra-body regions for subsequent external trajectory initialization. Specifically, the CT volume was thresholded from -250 to 3000 HU and the largest 3D connected component was isolated to eliminate background structures. This component was refined via slice-wise hole filling, removal of minor 2D artifacts, and a 2D morphological opening with a 5-mm radius. Finally, the original CT image providing validation intensities, the IRNet-predicted semantic mask providing candidate voxels, and the morphological body mask defining the external boundary were passed as integrated inputs to the downstream structure-aware digitization pipeline.

\subsubsection{Catheter reconstruction}\label{sec:catheter-reconstruction}

\begin{figure}[htbp]
\centering
\includegraphics[width=0.95\textwidth]{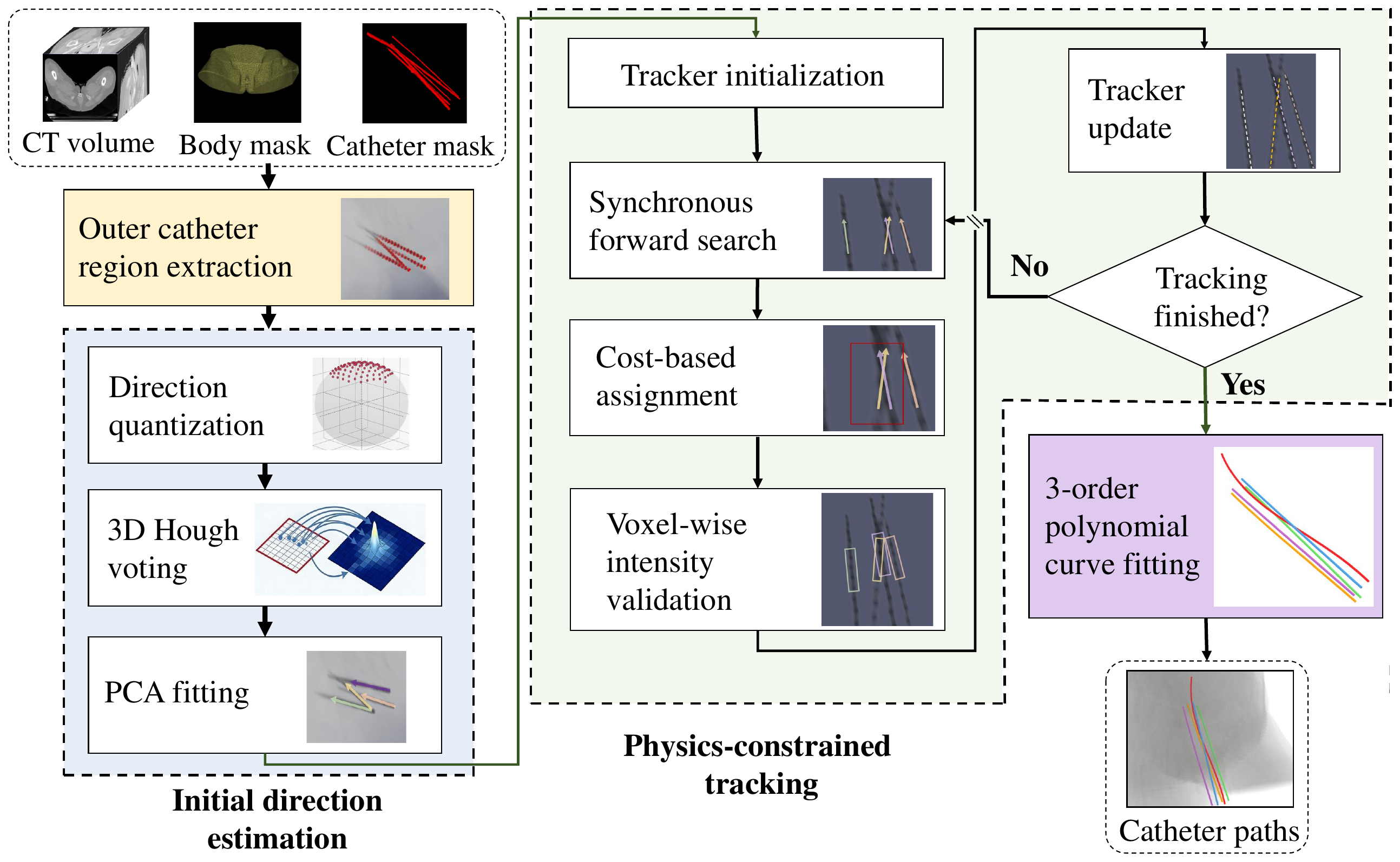}
\caption{Flowchart of the structure-aware trajectory reconstruction pipeline, detailing the sequential stages of external axis initialization via direction-constrained 3D Hough transform, synchronous physics-constrained inward tracking, and third-order polynomial regularization.}\label{fig:reconstruction}
\end{figure}

Although semantic segmentation isolates foreground catheter voxels, resolving them into individual trajectories remains challenging in dense and adherent configurations. We therefore developed a structure-aware reconstruction module motivated by catheter insertion biomechanics. The external catheter segment, which is not embedded in tissue, remains approximately straight and provides a reliable initial direction. In contrast, the internal segment may undergo gradual deflection because of heterogeneous needle--tissue interactions and anatomical interfaces, with the extent of deflection also affected by instrument stiffness and insertion conditions \cite{misraObservationsModelsNeedle2009,wangBevelTipNeedle2024,mcgillEffectsInsertionSpeed2012}. Its finite flexural rigidity nevertheless favors centerline continuity and gradual directional changes. Leveraging these physical priors, our pipeline first extracts reliable initial directions from the undeformed external segments. These extracted directions subsequently guide a synchronous, physics-constrained inward tracking process, which effectively recovers bent internal trajectories while minimizing ambiguity between neighboring catheters, as illustrated in Fig.~\ref{fig:reconstruction}.

\noindent\textbf{Initial direction estimation.} The initialization step aims to estimate a reliable coarse axis for each catheter prior to inward tracking. By restricting this estimation to the external catheter region, we avoid ambiguities introduced by internal tissue bending and catheter adhesion.

Let $\mathcal{P}=\{p_i\in\mathbb{R}^3,\ i=1,\ldots,N\}$ denote the segmented catheter voxels in physical coordinates, and let $B$ denote the body mask. The outer catheter voxel set was defined as

\begin{equation}
P_{\mathrm{out}}=\{p_i\in\mathcal{P}\mid p_i\notin B\}.
\end{equation}

To detect the dominant external catheter axes from $P_{\mathrm{out}}$, we applied an iterative 3D Hough transform based on Dalitz et al.\cite{dalitzIterativeHoughTransform2017}, introducing two application-specific modifications: direction-constrained Hough voting and CT-intensity-weighted axis refinement.

First, given that implanted catheters are predominantly oriented cranio-caudally, the candidate direction set was restricted to a 45$^\circ$ cone around the predefined insertion axis:

\begin{equation}
\mathcal{D}=\{b_j\mid \|b_j\|_2=1,\ \arccos(|b_j^T e_{cc}|)\le45^\circ\},
\end{equation}

where $e_{cc}$ denotes the cranio-caudal axis, and $b_j$ denotes the $j$-th discretized unit direction. This constrained sampling filters out anatomically implausible orientations, stabilizing axis detection in dense configurations.

Second, the inlier voxels of each extracted axis were refined via CT-intensity-weighted principal component analysis (PCA). For the $i$-th inlier voxel $p_i$ with intensity $I_i$, its weight was defined using min-max normalization within the segmented mask:

\begin{equation}
w_i=\frac{I_i-I_{\min}}{I_{\max}-I_{\min}},
\end{equation}

where $I_{\min}$ and $I_{\max}$ are the minimum and maximum CT intensities within the catheter mask. The weighted centroid $\mu$ and covariance matrix $\Sigma$ were computed as

\begin{equation}
\mu=\frac{\sum_i w_i p_i}{\sum_i w_i},\quad
\Sigma=\frac{\sum_i w_i(p_i-\mu)(p_i-\mu)^T}{\sum_i w_i}.
\end{equation}

The principal eigenvector of $\Sigma$ served as the refined initial direction. This weighting strategy prioritizes high-density metallic voxels, mitigating the impact of weak peripheral responses or segmentation noise. Together, the direction-constrained Hough transform and intensity-weighted PCA provided robust initial axes for the subsequent tracking procedure.

\noindent\textbf{Trajectory tracking and fitting.} While the initial directions enable a coarse separation of the external catheter portions, they are insufficient for recovering the full trajectories, as the internal segments often bend gradually and deviate from strict linearity. To address this limitation, we formulated and solved a synchronous constrained tracking problem. This approach assumes that each implanted catheter advances inward as a continuous, semi-rigid structure, characterized by limited lateral displacement and smooth directional transitions.

During initialization, external catheter voxels are assigned to their nearest detected axis. A tracker is initialized on the innermost point of the axis of each cluster, oriented inward. From this point, all trackers are advanced synchronously, iteration by iteration. This synchronous progression is crucial in dense configurations to prevent any single tracker from preemptively absorbing voxels that belong to adjacent, adhering catheters.

\noindent\textbf{Phase 1: Independent Candidate Search.} In each iteration, every active tracker $m$ with current position $x_m$ and direction $d_m$ independently identifies a pool of potential foreground voxels within a localized, forward-looking search volume. For a given voxel $p$ with displacement vector $\Delta p=p-x_m$, the forward distance $f_m^p$, radial distance $r_m^p$, and local directional consistency $\cos\theta_m(p)$ are computed as

\begin{equation}
f_m^p=\Delta p\cdot d_m,\quad
r_m^p=\|\Delta p-f_m^p d_m\|_2,\quad
\cos\theta_m(p)=\frac{f_m^p}{\|\Delta p\|_2}.
\end{equation}

A voxel was retained as a valid candidate only when it satisfies $f_m^p>0,\ f_m^p\le f_{\max},\ r_m^p\le r_{\max}$ and $\cos\theta_m(p)>\tau_{\theta}$. Clinically, this forward-looking search region reflects the limited lateral mobility of an implanted catheter, which is expected to continue locally along its current insertion direction rather than shift abruptly to a neighboring structure.

\noindent\textbf{Phase 2: Synchronous Cost-Based Assignment.} Because the search volumes of neighboring trackers may overlap, a single foreground voxel might satisfy the candidate criteria for multiple trackers simultaneously. To resolve these ambiguities globally, a cost-based assignment strategy is applied. The tracking cost for assigning voxel $p$ to tracker $m$ penalizes both lateral deviation and directional inconsistency:

\begin{equation}
C_m^p=r_m^p+\lambda\left(1-\cos\theta_m(p)\right).
\end{equation}

Each candidate voxel is exclusively assigned to the tracker that yields the minimum cost $C_m^p$. Consequently, candidate voxels that remain close to the current centerline and follow the estimated insertion direction are preferentially accepted, whereas abrupt lateral jumps toward adjacent adherent catheters are strongly penalized.

\noindent\textbf{Phase 3: Intensity Validation and State Update.} Following voxel assignment, each tracker calculates a tentative next centroid from its newly acquired voxels. Before confirming the move, a CT-intensity continuity validation step prevents the tracker from erroneously snapping to a neighboring catheter or disconnected high-density artifact. The line segment connecting the current position $x_m$ to the tentative centroid is sampled in the original CT image; the update is accepted only if the sampled intensities remain consistent with the estimated metallic-catheter intensity range.

Once validated, the tracking direction is adaptively updated using an inertia coefficient $\alpha\in[0,1)$ and the unit vector $u_m$ toward the validated centroid:

\begin{equation}
d_m^{\mathrm{new}}=\frac{(1-\alpha)u_m+\alpha d_m^{\mathrm{old}}}{\|(1-\alpha)u_m+\alpha d_m^{\mathrm{old}}\|_2}.
\end{equation}

The inertia term represents the finite flexural rigidity of the implanted catheter: gradual deflection caused by tissue resistance is permitted, whereas abrupt directional changes arising from segmentation noise or attraction to an adjacent catheter are suppressed. A tracker terminates when it fails to secure any validated candidate voxels within the forward-looking neighborhood for a maximum of $N_{\mathrm{fail}}$, empirically set to 3, consecutive iterations.

Finally, the discrete point sequence of each catheter is regularized by third-order polynomial curve fitting. A global reassignment step is then performed by computing the minimum Euclidean distance from each unassigned foreground voxel to the fitted curves, successfully achieving robust separation and digitization of multiple closely adjacent or adhesion catheters.

\subsection{Implementation Details}\label{sec:implementation-details}

\subsubsection{Comparison methods}\label{sec:comparison-methods}

The proposed framework was evaluated across two distinct tasks: semantic catheter segmentation and catheter digitization. For segmentation, the proposed IRNet was benchmarked against a 2D nnU-Net, used as a slice-wise baseline, and a 3D full-resolution nnU-Net, used to assess the contribution of volumetric context\cite{isenseeNnUNetSelfconfiguringMethod2021}. All models were evaluated using identical patient-level five-fold test splits and standardized evaluation protocols.

Three training regimes were considered. In the synthetic-only regime, each model was trained from random initialization using 123 synthetic CT volumes. In the fully supervised clinical regime, each model was trained from random initialization using all annotated fractions from the training patients in each fold, without synthetic pretraining. This setting served as a fully supervised reference. To evaluate annotation efficiency, IRNet was additionally pretrained on the synthetic dataset and fine-tuned using nested subsets of 1, 5, 10, or 20 annotated clinical fractions selected from the training patients in each fold, where one shot corresponded to one clinical fraction. All models were evaluated on the same held-out patients in each fold.

For catheter digitization, the evaluation was organized into two complementary settings. Jung et al.\cite{jungDeeplearningAssistedAutomatic2019} and Quetin et al.\cite{quetin2025automatic} were selected as baselines because both explicitly address catheter adhesion and therefore provide directly relevant comparators for the principal reconstruction challenge investigated in this study. First, intrinsic reconstruction performance was assessed by applying the proposed method and the two adhesion-aware baselines to ground-truth semantic masks, thereby isolating reconstruction performance from upstream segmentation errors. Second, end-to-end pipeline performance was evaluated by applying the proposed reconstruction method to masks generated by two IRNet configurations on held-out patients. The fully supervised IRNet, trained using all annotated fractions from the training patients in each fold, served as the upper-bound reference. The data-efficient configuration used an IRNet pretrained on synthetic data and fine-tuned with only 10 annotated clinical fractions. Both evaluation settings were analyzed in the whole cohort and the predefined catheter-adhesion subgroup. A treatment fraction was assigned to the catheter-adhesion subgroup when two or more catheters were in direct contact or exhibited overlapping foreground regions across consecutive axial CT slices, resulting in locally merged semantic catheter regions and ambiguous instance separation. Based on this criterion, 83 of the 203 treatment fractions were classified as catheter-adhesion cases.

\subsubsection{Training and inference}\label{sec:training-and-inference}

Model training and inference were performed on a workstation equipped with an Intel Xeon Gold 6126 CPU at 2.60 GHz and an NVIDIA TITAN RTX GPU with 24 GB of memory. Synthetic CT generation was implemented using Elastix 5.2.0\cite{5338015} and SimpleITK\cite{lowekampDesignSimpleITK2013} for rigid registration between the target pelvic CT and the reference CT. Forty-one pelvic CT cases from the CTPelvic1K\cite{liuDeepLearningSegment2021} dataset were used as anatomical backgrounds. For each case, three independent augmented catheter configurations were generated, resulting in a total of 123 synthetic 3D CT volumes for model training. Metal artifact synthesis was performed using an established simulation method\cite{zhangConvolutionalNeuralNetwork2018}.

For segmentation model training, the 2D nnU-Net and the proposed IRNet were trained for 100 epochs, after which stable performance was observed. The training time for each 2D model was approximately 1.5 h. The 3D full-resolution nnU-Net was trained for a fixed 150 epochs and required approximately 4 h. All models were trained using the exact same simulated dataset to ensure a fair comparison. To establish the fully supervised clinical reference, identical network architectures and training procedures were implemented within the same patient-level five-fold cross-validation framework, using only the training patients in each fold for model optimization. For the data-efficient setting, IRNet was pretrained on 123 synthetic CT volumes and then fine-tuned under a 10-shot setting using annotated clinical samples selected from the training patients in each fold. The fine-tuned IRNet was applied to the corresponding held-out patients to generate segmentation masks for end-to-end digitization. Thus, no clinical fraction from the held-out patients contributed to either fine-tuning or fully supervised training in the same fold. During fine-tuning, the heatmap-aware loss was preserved under a full-network optimization scheme. To mitigate the risk of overfitting on the limited clinical dataset, training was restricted to 100 epochs. Furthermore, a differential low-learning-rate strategy was adopted: the learning rates for the encoder and the decoder, including task-specific heads, were conservatively set to $1\times10^{-5}$ and $5\times10^{-5}$, respectively, coupled with a weight decay of $1\times10^{-4}$. This configuration was designed to evaluate whether minimal clinical annotations can effectively bridge the synthetic-to-real domain gap while maintaining the annotation efficiency of the framework.

To systematically evaluate the annotation efficiency of the synthetic-to-real strategy, we assessed model performance under 1-, 5-, 10-, and 20-shot fine-tuning settings. These subsets were nested to ensure a fair comparison across different annotation budgets. To preclude data leakage, all fine-tuning subsets were strictly isolated from the independent evaluation cohort. The pretrained IRNet checkpoint, optimization scheme, and inference protocols were kept constant across all configurations to isolate the impact of annotation volume on segmentation accuracy.

For catheter digitization, the Hough discretization step and icosahedral subdivision level were set to 2.0 mm and 4, respectively. During constrained tracking, the maximum forward distance $f_{\max}$, radial search radius $r_{\max}$, and directional threshold $\tau_{\theta}$ were set to 8.0 mm, 2 mm, and 0.4, respectively. The cost-balancing weight $\lambda$, regulating the trade-off between spatial proximity and physical insertion consistency, and the physical direction-inertia coefficient $\alpha$ were empirically set to 5.0 and 0.5. The metal-intensity threshold for continuity validation was defined as half of the median CT intensity within the segmented catheter mask.

\subsubsection{Evaluation metrics}\label{sec:evaluation-metrics}

Dice similarity coefficient and the 95th percentile Hausdorff distance (HD95) were used to evaluate the segmentation performance. For catheter digitization, predicted ($P_i$) and reference catheters ($G_j$) were matched using a trajectory-based assignment strategy\cite{quetin2025automatic}. A one-to-one assignment was obtained by using the Hungarian algorithm to minimize the matching cost matrix:

\begin{equation}
D(G_j,P_i)=\frac{1}{|G_j|}\sum_{g\in G_j}\min_{p\in P_i}\|g-p\|_2.
\end{equation}

An assigned pair was considered a true positive (TP) only if the shaft matching cost $D(G_j,P_i)\le3.0\ \mathrm{mm}$ and the tip distance $\|t(P_i)-t(G_j)\|_2\le6.0\ \mathrm{mm}$. Unmatched predicted and reference catheters were recorded as false positives and false negatives, respectively, to compute precision, recall, and the F1-score\cite{zhangMultineedleLocalizationAttention2020,zhangAutomaticMultineedleLocalization2020}.

For each TP pair, tip digitization error ($E_{\mathrm{tip}}$) and shaft digitization error ($E_{\mathrm{shaft}}$) were evaluated to assess geometric agreement. $E_{\mathrm{tip}}$ was defined as the Euclidean distance between the corresponding endpoints. $E_{\mathrm{shaft}}$ was defined as the bidirectional mean closest-point distance between the centerlines:

\begin{equation}
E_{\mathrm{shaft}}=\frac{1}{2}\left(\frac{1}{|P|}\sum_{p\in P}\min_{g\in G}\|p-g\|_2+\frac{1}{|G|}\sum_{g\in G}\min_{p\in P}\|g-p\|_2\right).
\end{equation}

All metrics were computed per case and averaged across the test set. To evaluate the robustness of the structure-aware reconstruction module, a local sensitivity analysis was performed by varying one parameter at a time while keeping the remaining parameters fixed at their baseline settings. The tested values were 6--10 mm for $f_{\max}$, 1.5--2.5 mm for $r_{\max}$, 0.2--0.6 for $\tau_{\theta}$, 2.5--7.5 mm for $\lambda$, and 0.3--0.7 for $\alpha$. For the sensitivity analysis, patient-level cluster-bootstrap 95\% confidence intervals were estimated for the pooled F1 score, $E_{\mathrm{shaft}}$, and $E_{\mathrm{tip}}$ using 5,000 bootstrap resamples. In each resample, patients were sampled with replacement, and all treatment fractions from each selected patient were retained to account for within-patient clustering.

\section{Results}\label{sec:results}

\subsection{Evaluation of the segmentation results}\label{sec:evaluation-of-the-segmentation-results}

\subsubsection{Qualitative assessment of synthetic CT images}\label{sec:qualitative-synthetic-ct}

\begin{figure}[htbp]
\centering
\includegraphics[width=0.95\textwidth]{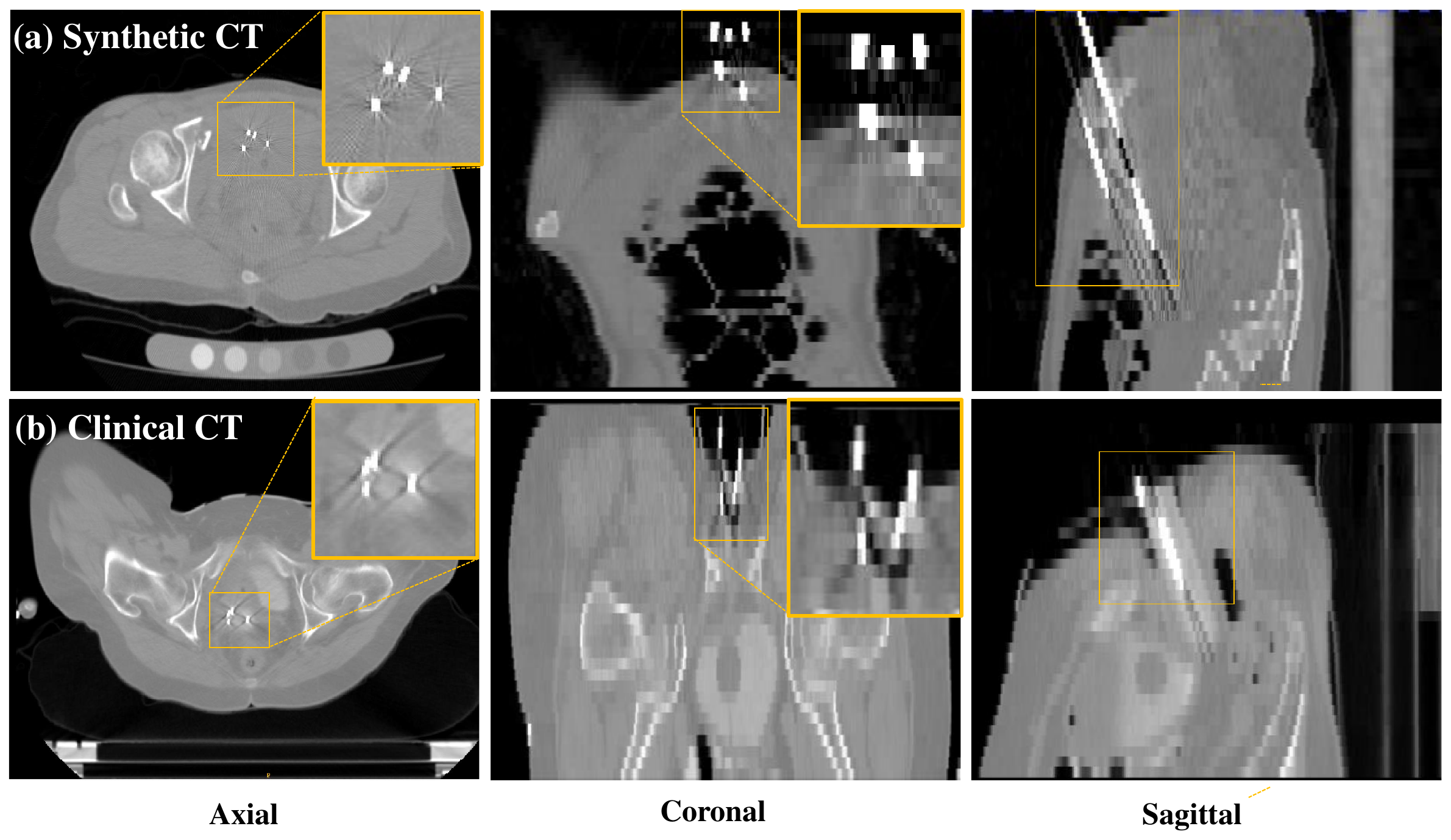}
\caption{Qualitative comparison between simulated and clinical post-implant CT images. The simulated CT image reproduces the high-intensity catheter appearance and surrounding metal-induced artifacts observed in clinical CT. Zoomed-in regions show the catheter cores and local artifact patterns.}\label{fig:sim-clinical}
\end{figure}

Before reporting the segmentation metrics, we first performed a qualitative comparison between simulated and clinical post-implant CT images. As shown in Fig.~\ref{fig:sim-clinical}, the simulated CT images reproduced the key catheter-related appearances observed in clinical CT, including high-intensity catheter cores, local blooming, partial-volume blurring, and surrounding streak artifacts. Although the simulated artifacts appeared slightly stronger and more regular than those in clinical scans, the overall catheter contrast and artifact morphology were visually consistent with the clinical images. This observation supports the use of simulated CT images as task-relevant supervision for catheter segmentation.

\subsubsection{Quantitative assessment of segmentation methods}\label{sec:quantitative-segmentation}

Table~\ref{tab:segmentation} summarizes the quantitative segmentation performance. When trained exclusively on the synthetic source domain, IRNet achieved strong voxel-wise overlap with a Dice coefficient of 0.937 $\pm$ 0.163, outperforming baseline architectures and validating the utility of simulated artifacts for learning localized metallic signatures. However, zero-shot generalization to clinical CTs revealed geometric vulnerabilities across all synthetic-only models, evidenced by extreme HD95 standard deviations, such as 4.34 $\pm$ 24.82 mm for IRNet and 9.40 $\pm$ 35.07 mm for 2D nnU-Net. These skewed variances are driven by severe case-level outliers, where models erroneously over-segmented distant, high-density anatomical structures such as cortical bone that were insufficiently represented in the synthetic background.

While fully supervised clinical training systematically mitigated these domain-shift errors and improved Dice scores, standard baseline architectures still lacked geometric stability, as reflected by a clinical 3D nnU-Net HD95 of 14.00 $\pm$ 48.09 mm. In contrast, the clinically trained IRNet achieved the lowest overall HD95 of 0.67 $\pm$ 1.12 mm. This result indicates that the proposed implant-region-aware mechanism improves spatial compactness; by explicitly modulating features to suppress anatomically implausible background predictions, IRNet effectively reduces the geometric outliers that standard architectures struggle to handle.

\subsubsection{Effect of the clinical annotation budget}\label{sec:annotation-budget}

To determine the minimal clinical annotation budget required to bridge the aforementioned synthetic-to-real domain gap, we evaluated the pretrained IRNet under progressive few-shot fine-tuning regimes of 1, 5, 10, and 20 fractions. Under extreme low-shot conditions of 1 and 5 shots, IRNet maintained moderate voxel-wise overlap, with Dice coefficients of 0.900 $\pm$ 0.064 and 0.914 $\pm$ 0.043, respectively, yet geometric reliability remained compromised, with HD95 values of 6.433 $\pm$ 11.829 mm and 3.324 $\pm$ 6.524 mm.

Expanding the fine-tuning subset to 10 annotated clinical fractions reduced the influence of these geometric outliers, lowering the HD95 to 0.853 $\pm$ 0.362 mm while stabilizing the Dice coefficient at 0.927 $\pm$ 0.034. Increasing the budget further to 20 fractions yielded only marginal gains in the Dice score of 0.935 $\pm$ 0.052 and a mean HD95 of 0.951 $\pm$ 0.257 mm, which did not further improve over the 10-shot setting despite slightly reduced HD95 variability. These findings demonstrate that limited fine-tuning predominantly resolves spatial ambiguities and corrects geometric outliers, rather than substantially altering basic volumetric overlap. Consequently, the 10-shot regime provides a favorable trade-off between segmentation performance and annotation efficiency.

\begin{table}[htbp]
\caption{Segmentation performance across network architectures, training domains, and clinical annotation budgets. Values are reported as mean $\pm$ standard deviation across held-out test cases. One shot denotes one annotated clinical fraction. Dice, Dice similarity coefficient; HD95, 95th-percentile Hausdorff distance.}
\label{tab:segmentation}%
\begin{tabular}{@{}p{0.24\textwidth}p{0.31\textwidth}p{0.17\textwidth}p{0.17\textwidth}@{}}
\toprule
Method & Training setting & Dice & HD95 (mm)\\
\midrule
2D nnU-Net & Synthetic only & 0.928 $\pm$ 0.158 & 9.40 $\pm$ 35.07\\
3D nnU-Net & Synthetic only & 0.900 $\pm$ 0.145 & 15.72 $\pm$ 53.23\\
IRNet & Synthetic only & 0.937 $\pm$ 0.163 & 4.34 $\pm$ 24.82\\
\midrule
2D nnU-Net & Fully supervised & 0.961 $\pm$ 0.034 & 0.84 $\pm$ 2.00\\
3D nnU-Net & Fully supervised & 0.942 $\pm$ 0.040 & 14.00 $\pm$ 48.09\\
IRNet & Fully supervised & 0.958 $\pm$ 0.037 & 0.67 $\pm$ 1.12\\
\midrule
IRNet & 1-shot fine-tuning & 0.900 $\pm$ 0.064 & 6.433 $\pm$ 11.829\\
IRNet & 5-shot fine-tuning & 0.914 $\pm$ 0.043 & 3.324 $\pm$ 6.524\\
IRNet & 10-shot fine-tuning & 0.927 $\pm$ 0.034 & 0.853 $\pm$ 0.362\\
IRNet & 20-shot fine-tuning & 0.935 $\pm$ 0.052 & 0.951 $\pm$ 0.257\\
\botrule
\end{tabular}
\end{table}

\subsubsection{Qualitative analysis of IRNet predictions}\label{sec:qualitative-irnet}

\begin{figure}[htbp]
\centering
\includegraphics[width=1\textwidth]{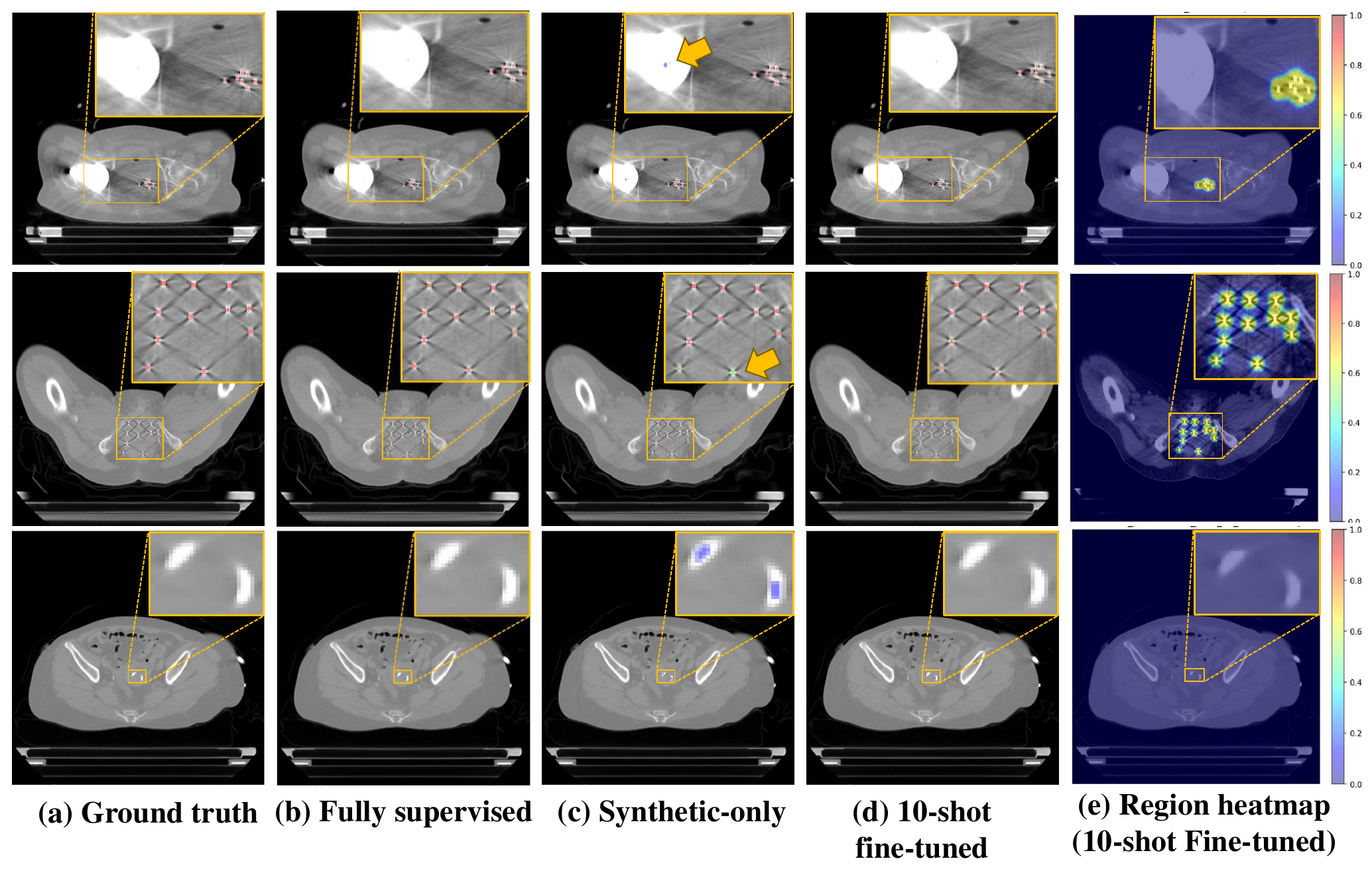}
\caption{Qualitative assessment of catheter segmentation performance. Regions of interest (yellow bounding boxes) are magnified to illustrate local boundary delineations. Accurately predicted pixels are colored red, while false-negative omissions and false-positive over-segmentations are highlighted in green and blue, respectively.}\label{fig:segmentation}
\end{figure}

Figure~\ref{fig:segmentation} visually corroborates these quantitative improvements. Compared with the synthetic-only model, fine-tuning reduced distant false-positive predictions while preserving catheter continuity. The 10-shot model produced masks visually similar to those of the fully supervised IRNet. The corresponding predicted implant-region heatmaps confirm that IRNet provides robust, highly concentrated spatial attention within the implant region, avoiding the erroneous responses seen in the baselines.

\subsection{Evaluation of the reconstruction results}\label{sec:evaluation-of-the-reconstruction-results}

\subsubsection{End-to-end reconstruction performance}\label{sec}

The proposed framework enables fully automated end-to-end reconstruction from clinical CT images to instance-level catheter trajectories, without requiring manually delineated masks during inference. As shown in Part A of Table~\ref{tab:digitization}, both the fully supervised and 10-shot adapted pipelines achieved robust reconstruction performance in the whole cohort and in the catheter-adhesion subgroup. The limited performance difference between the two training settings indicates that synthetic pretraining followed by few-shot clinical adaptation preserves most of the reconstruction capability obtained with full clinical supervision. This finding supports the feasibility of deploying the proposed framework under limited clinical annotation availability.

To benchmark the proposed reconstruction strategy against existing methods, we additionally conducted a controlled comparison in which ground-truth semantic masks were provided as a common input to all reconstruction algorithms. This setting eliminates the influence of differences in upstream mask prediction and enables a fair comparison of the intrinsic reconstruction capabilities of the different methods. As shown in Part B of Table~\ref{tab:digitization}, the proposed method consistently outperformed the methods adapted from Jung et al.\cite{jungDeeplearningAssistedAutomatic2019} and Quetin et al.\cite{quetin2025automatic}. Its advantage was maintained in the catheter-adhesion subgroup, where separating closely adjacent trajectories and preserving individual catheter identities are particularly challenging. These results demonstrate that the proposed reconstruction strategy provides substantial advantages over existing methods, especially in complex catheter configurations.

Comparison between Parts A and B further demonstrates that the catheter-level performance of the complete pipeline remained close to that obtained with ground-truth masks. Although predicted masks introduced larger pointwise shaft and tip errors, their influence on catheter detection and trajectory recovery was relatively limited. This suggests that successful reconstruction depends primarily on preserving catheter topology, centerline continuity, and directional information rather than achieving exact voxel-level boundary delineation. The proposed reconstruction module can therefore tolerate moderate upstream prediction errors and recover coherent catheter trajectories from automatically generated masks.

\begin{table}[!htbp]
\caption{Catheter digitization performance in the whole cohort and catheter-adhesion subgroup.}\label{tab:digitization}%
\centering
\setlength{\tabcolsep}{5pt}%
\renewcommand{\arraystretch}{1.12}%
\small
\begin{tabular*}{\textwidth}{@{\extracolsep{\fill}}p{0.22\textwidth}lcc@{}}
\multicolumn{4}{@{}l}{\textit{Part A. End-to-end reconstruction performance}}\\[3pt]
\toprule
Cohort & Metric & \makecell{Fully supervised \\ IRNet} & \makecell{10-shot fine-tuned \\ IRNet}\\
\midrule
\multirow{5}{0.22\textwidth}{Whole cohort}
& $E_{\mathrm{shaft}}$ (mm) & 0.311 $\pm$ 0.269 & 0.334 $\pm$ 0.367\\
& $E_{\mathrm{tip}}$ (mm)   & 0.878 $\pm$ 0.585 & 0.896 $\pm$ 0.680\\
& Recall                    & 0.911 $\pm$ 0.177 & 0.894 $\pm$ 0.177\\
& Precision                 & 0.907 $\pm$ 0.182 & 0.890 $\pm$ 0.180\\
& F1                        & 0.909 $\pm$ 0.179 & 0.891 $\pm$ 0.178\\
\addlinespace
\multirow{5}{0.22\textwidth}{Catheter-adhesion subgroup}
& $E_{\mathrm{shaft}}$ (mm) & 0.352 $\pm$ 0.336 & 0.404 $\pm$ 0.476\\
& $E_{\mathrm{tip}}$ (mm)   & 0.875 $\pm$ 0.453 & 0.876 $\pm$ 0.760\\
& Recall                    & 0.847 $\pm$ 0.226 & 0.844 $\pm$ 0.191\\
& Precision                 & 0.841 $\pm$ 0.229 & 0.842 $\pm$ 0.191\\
& F1                        & 0.844 $\pm$ 0.227 & 0.843 $\pm$ 0.190\\
\botrule
\end{tabular*}
\vspace{8pt}
\begin{tabular*}{\textwidth}{@{\extracolsep{\fill}}p{0.22\textwidth}lccc@{}}
\multicolumn{5}{@{}l}{\textit{Part B. Comparison of reconstruction methods using ground-truth masks}}\\[3pt]
\toprule
Cohort & Metric & Jung et al. & Quetin et al. & Proposed\\
\midrule
\multirow{5}{0.22\textwidth}{Whole cohort}
& $E_{\mathrm{shaft}}$ (mm) & 0.226 $\pm$ 0.475 & 0.185 $\pm$ 0.441 & 0.067 $\pm$ 0.203\\
& $E_{\mathrm{tip}}$ (mm)   & 0.087 $\pm$ 0.351 & 0.181 $\pm$ 0.529 & 0.060 $\pm$ 0.226\\
& Recall                    & 0.798 $\pm$ 0.279 & 0.795 $\pm$ 0.289 & 0.923 $\pm$ 0.157\\
& Precision                 & 0.798 $\pm$ 0.279 & 0.795 $\pm$ 0.289 & 0.919 $\pm$ 0.161\\
& F1                        & 0.798 $\pm$ 0.279 & 0.795 $\pm$ 0.289 & 0.921 $\pm$ 0.159\\
\addlinespace
\multirow{5}{0.22\textwidth}{Catheter-adhesion subgroup}
& $E_{\mathrm{shaft}}$ (mm) & 0.307 $\pm$ 0.560 & 0.339 $\pm$ 0.594 & 0.086 $\pm$ 0.207\\
& $E_{\mathrm{tip}}$ (mm)   & 0.126 $\pm$ 0.431 & 0.249 $\pm$ 0.553 & 0.099 $\pm$ 0.300\\
& Recall                    & 0.740 $\pm$ 0.303 & 0.699 $\pm$ 0.299 & 0.877 $\pm$ 0.180\\
& Precision                 & 0.740 $\pm$ 0.303 & 0.699 $\pm$ 0.299 & 0.874 $\pm$ 0.184\\
& F1                        & 0.740 $\pm$ 0.303 & 0.699 $\pm$ 0.299 & 0.875 $\pm$ 0.182\\
\botrule
\end{tabular*}
\vspace{5pt}
\begin{minipage}{\textwidth}
\footnotesize
\textit{Note:} The identical recall, precision, and F1 values reported for the methods adapted from Jung et al. and Quetin et al. arise from their count-constrained reconstruction strategies, in which the number of reconstructed catheters is fixed according to the known catheter count. Consequently, the numbers of false-positive and false-negative trajectories are equal, resulting in numerically identical precision, recall, and F1 values.
\end{minipage}
\end{table}

\subsubsection{Robustness in the catheter-adhesion subgroup}\label{sec:adhesion-robustness}

Robustness was further evaluated in the predefined catheter-adhesion subgroup, which contains geometrically challenging cases in which adjacent catheter foreground regions may remain connected across multiple CT slices. As shown in Part A of Table~\ref{tab:digitization}, the fully supervised end-to-end pipeline achieved a shaft digitization error of 0.352 $\pm$ 0.336 mm, a tip digitization error of 0.875 $\pm$ 0.453 mm, and an F1 score of 0.844 $\pm$ 0.227. The corresponding values obtained with the 10-shot fine-tuned IRNet were 0.404 $\pm$ 0.476 mm, 0.876 $\pm$ 0.760 mm, and 0.843 $\pm$ 0.190, respectively. The close performance of the two settings indicates that synthetic pretraining followed by limited fine-tuning preserved most of the end-to-end reconstruction capability achieved with full clinical supervision, even in cases involving closely adjacent or adherent catheters.

Part B of Table~\ref{tab:digitization} compares the reconstruction methods under the same ground-truth mask input. The proposed method achieved a shaft digitization error of 0.086 $\pm$ 0.207 mm, a tip digitization error of 0.099 $\pm$ 0.300 mm, and an F1 score of 0.875 $\pm$ 0.182. These results remained superior to those of the methods adapted from Jung et al. and Quetin et al. This controlled comparison indicates that the proposed reconstruction strategy is more effective at separating closely adjacent trajectories and preserving catheter identity under adherent configurations. Together, the results from Parts A and B demonstrate that the proposed framework retains robust end-to-end performance in challenging catheter-adhesion cases, while its downstream reconstruction strategy provides an independent advantage over existing methods.

\begin{figure}[!htbp]
\centering
\includegraphics[width=\textwidth]{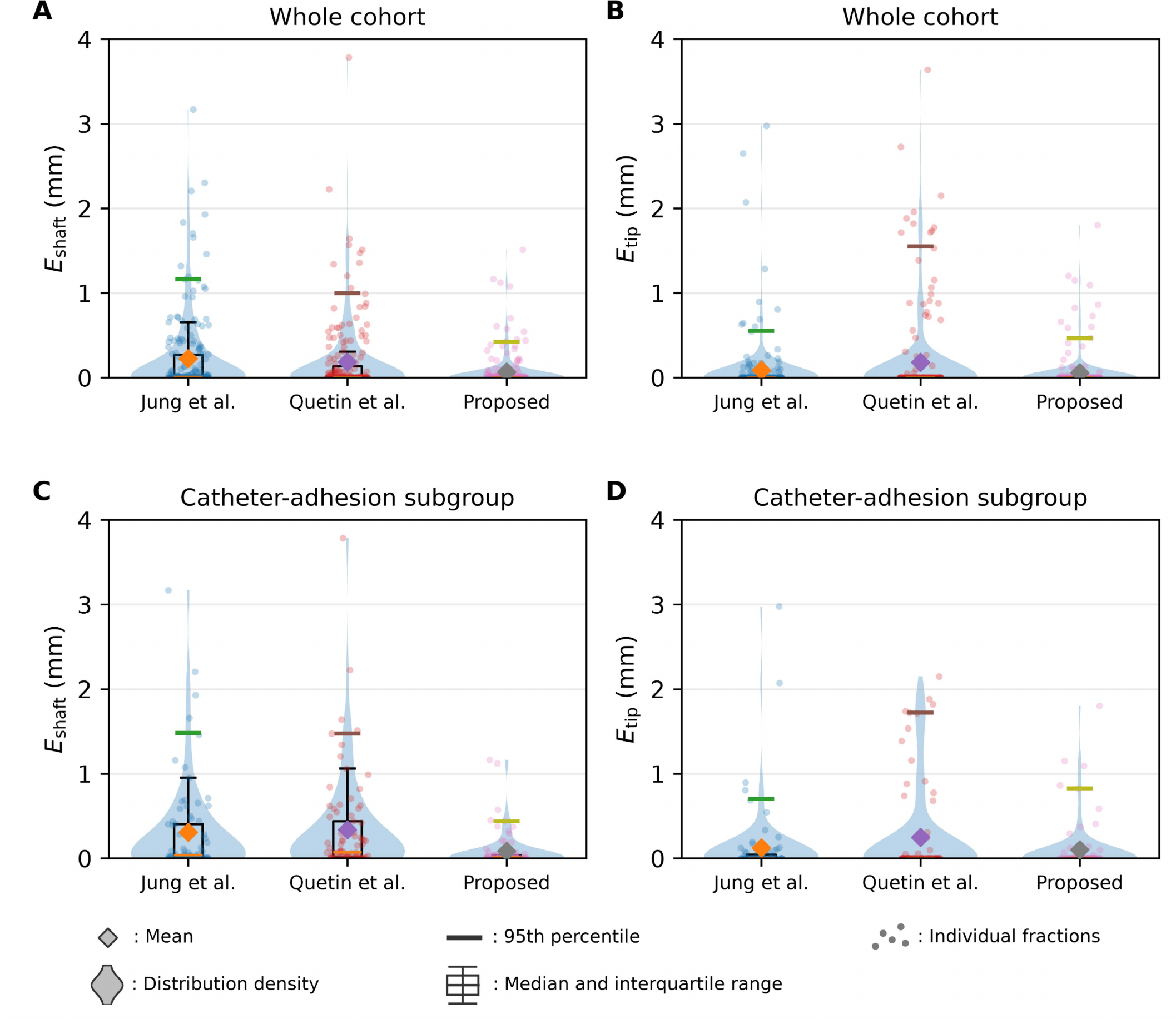}
\caption{Distributional comparison of catheter reconstruction errors using ground-truth semantic masks as the common input. Panels A and B show the shaft digitization error ($E_{\mathrm{shaft}}$) and tip digitization error ($E_{\mathrm{tip}}$), respectively, for the whole cohort. Panels C and D show the corresponding error distributions for the catheter-adhesion subgroup. Violin plots represent distribution density, individual points denote treatment fractions, boxes indicate the median and interquartile range, diamonds indicate the mean, and horizontal bars indicate the 95th percentile. Lower values indicate better reconstruction accuracy.}
\label{fig:error-distributions}
\end{figure}

Figure~\ref{fig:error-distributions} further characterizes the fraction-level distributions of shaft and tip digitization errors when all reconstruction methods were provided with the same ground-truth semantic masks. Across both the whole cohort and the catheter-adhesion subgroup, the proposed method produced error distributions that were more concentrated near zero, with lower central tendencies and reduced upper-tail errors than the comparison methods. The improvement was particularly evident for shaft digitization, indicating that the global initialization and synchronous physics-constrained tracking effectively suppress cumulative trajectory deviation. Although isolated outliers remained in complex adhesion cases, the consistently compact distributions suggest that the observed performance gains were maintained across treatment fractions rather than being driven by a small number of favorable cases.

\subsubsection{Qualitative comparison of reconstruction results}\label{sec:qualitative-reconstruction}

\begin{figure}[htbp]
\centering
\includegraphics[width=0.95\textwidth]{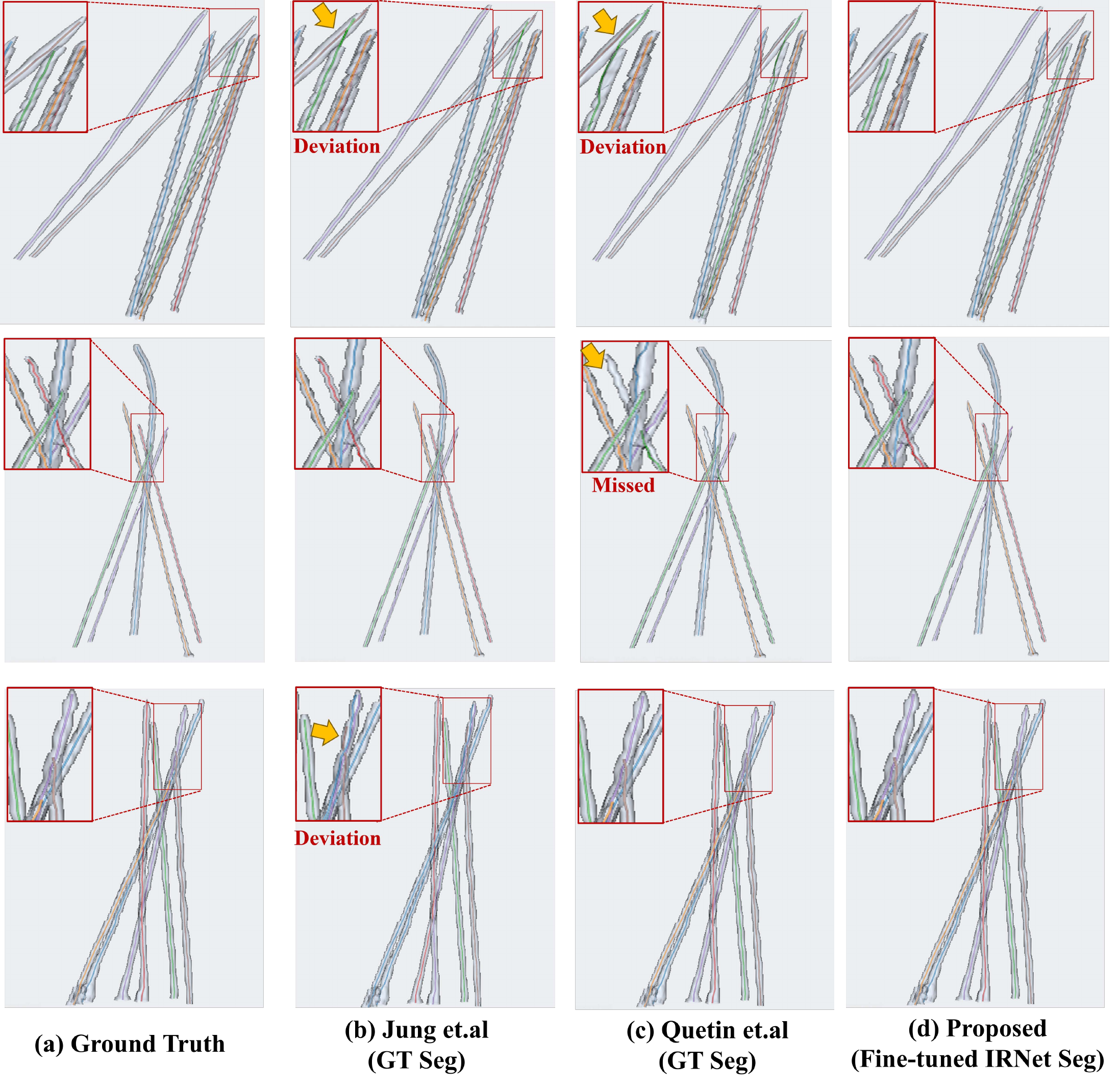}
\caption{Qualitative comparison of catheter reconstruction results across different methods. The reconstructed catheter masks are rendered in light gray, and the centerline of each catheter is overlaid in a distinct color for visual differentiation. Zoomed-in regions are highlighted by red boxes, while orange arrows indicate representative digitization errors, including trajectory deviations and missed catheters.}
\label{fig:reconstruction-results}
\end{figure}

Qualitative comparisons in Fig.~\ref{fig:reconstruction-results} illustrate the reconstruction results under representative challenging configurations. In tightly clustered or intersecting regions, the compared methods occasionally showed trajectory deviation or missed catheter segments, suggesting that local separation can become ambiguous when adjacent catheters are highly adherent. By using stable external catheter segments for 3D Hough-based initialization and then applying physics-constrained inward tracking, the proposed method generally preserved more continuous and anatomically plausible trajectories.

\subsubsection{Hyperparameter sensitivity analysis}
\label{sec:hyperparameter-sensitivity}

\begin{figure}[!htbp]
\centering
\makebox[\textwidth][c]{\includegraphics[width=1.12\textwidth]{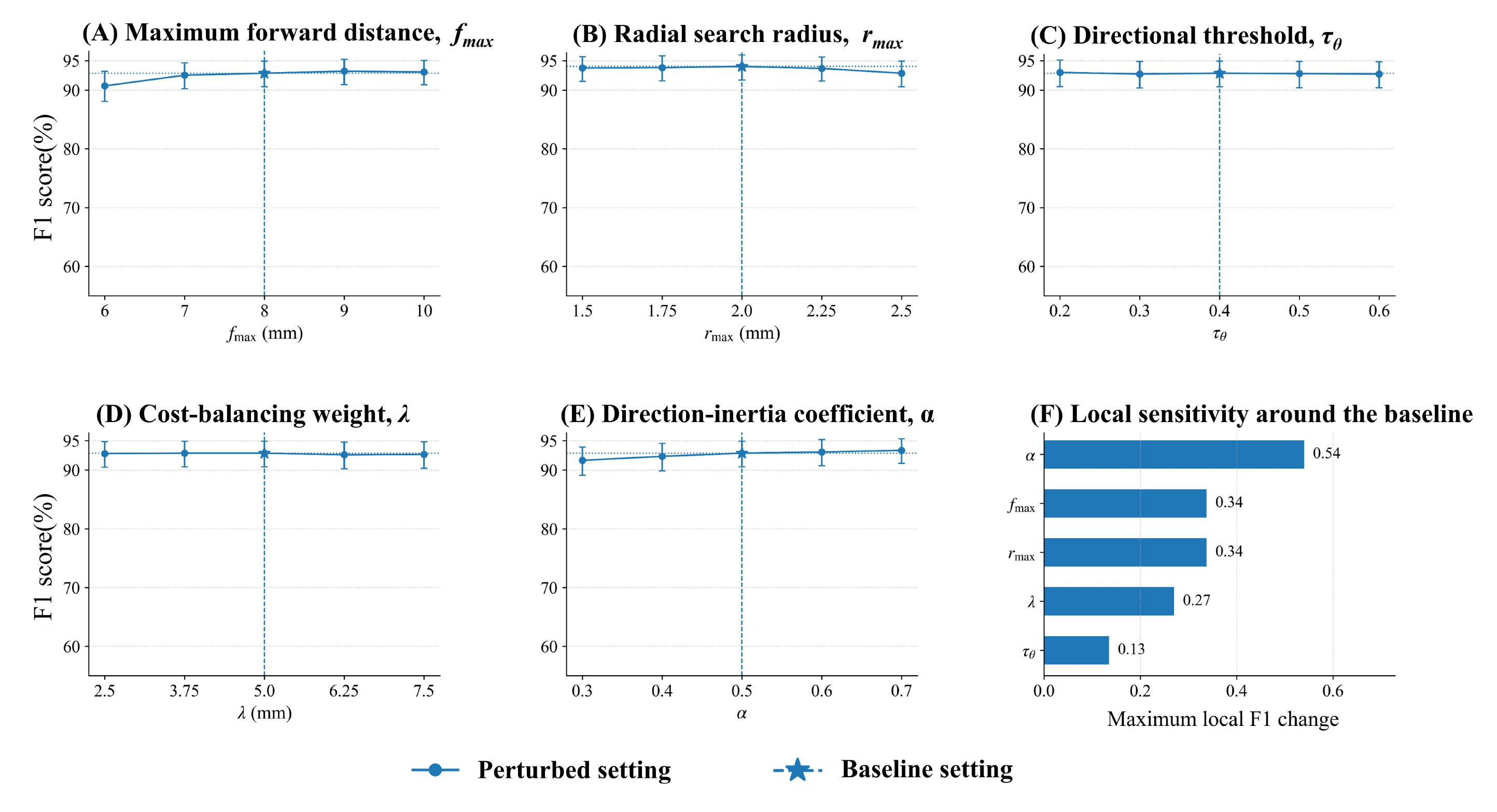}}
\caption{Local sensitivity analysis of the five principal tracking parameters. Panels A--E show the pooled F1 score with patient-level cluster-bootstrap 95\% confidence intervals when one parameter was varied while the remaining parameters were fixed at their baseline settings. Stars and vertical dashed lines indicate the baseline values. Panel F summarizes the maximum absolute F1 change at the two tested levels nearest to each baseline.}
\label{fig:hyperparameter-sensitivity}
\end{figure}

Figure~\ref{fig:hyperparameter-sensitivity} shows the local sensitivity of the F1 score to the five principal tracking parameters. The pooled F1 score remained above 90\% across all tested settings, with no abrupt performance deterioration around the baseline configuration. The maximum local F1 changes near the baseline were 0.34, 0.34, 0.13, 0.27, and 0.54 percentage points for $f_{\max}$, $r_{\max}$, $\tau_{\theta}$, $\lambda$, and $\alpha$, respectively. These results indicate that the selected parameter configuration lies within a stable operating region.

Consistent with the F1-based analysis, geometric accuracy remained stable across the tested parameter settings. Relative to the baseline configuration, the maximum increases in $E_{\mathrm{shaft}}$ and $E_{\mathrm{tip}}$ were 9.7\% and 16.8\%, respectively, while several parameter perturbations yielded comparable or lower errors. These results indicate that moderate variations in the tracking parameters did not materially degrade reconstruction accuracy.

\subsection{Computational efficiency}\label{sec:computational-efficiency}

To evaluate the clinical feasibility of the proposed framework, we assessed the computational efficiency of its two main stages: IRNet segmentation inference and post-segmentation catheter reconstruction. On average, the IRNet segmentation required 7.208 $\pm$ 3.238 seconds per case, or 0.143 $\pm$ 0.077 seconds per slice. For the downstream reconstruction stage, our method took 4.436 $\pm$ 1.903 seconds per case, compared to 3.927 $\pm$ 1.217 seconds and 2.544 $\pm$ 1.420 seconds for the methods proposed by Jung et al. and Quetin et al., respectively. Consequently, the complete end-to-end workflow achieved an average processing time of approximately 11.6 seconds per case. Although our digitization step incurs a marginal increase in runtime compared to the baselines, its execution time of less than 5 seconds remains highly efficient, confirming the overall framework's suitability for routine clinical integration.

From a theoretical perspective, the computational efficiency of our catheter reconstruction algorithm stems from its linear time complexity relative to the number of foreground voxels, $V$. Let $N$ represent the number of catheters and $K$ denote the number of discretized bins in the Hough parameter space. The two primary stages, the external 3D Hough transform and the subsequent inward tracking, require $O(V\cdot K)$ and $O(V\cdot N)$ operations, respectively, bounding the overall time complexity at $O(V(K+N))$. Because the upstream segmentation filters out the vast majority of the anatomical background, $V$ remains exceptionally small. This strictly constrained search space ensures the method scales gracefully, maintaining predictable execution times even in dense, multi-catheter configurations.

\subsection{Analysis of failure cases}\label{sec:analysis-of-failure-cases}

\begin{figure}[htbp]
\centering
\includegraphics[width=0.95\textwidth]{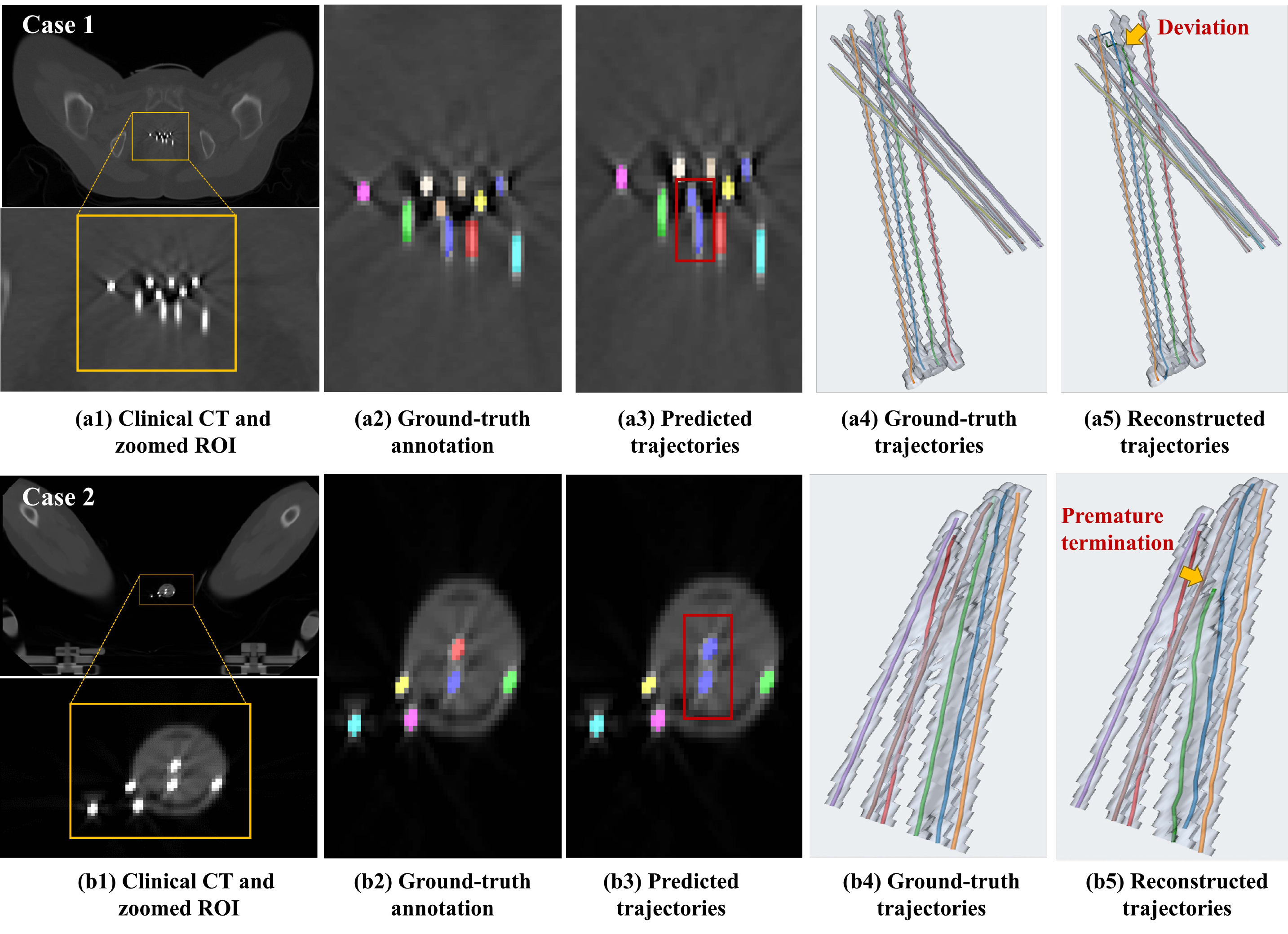}
\caption{Two representative failure cases of the proposed catheter localization method. From left to right, each row shows the axial CT image with a zoomed-in region of interest, the corresponding 2D localization results, and the 3D trajectory visualizations. In Case 1, two catheters exhibit an evident deviation from the expected trajectory. In Case 2, one reconstructed trajectory shows premature termination, ending before reaching the true endpoint. Red boxes and yellow arrows indicate the corresponding reconstruction errors.}\label{fig:failure-cases}
\end{figure}

Despite the overall robustness of the proposed framework, trajectory reconstruction remains challenging in highly adherent multi-catheter configurations. When several catheters remain in close contact or overlap across consecutive axial slices, their semantic foreground regions may merge, making it difficult to preserve individual catheter identities during inward tracking. Figure~\ref{fig:failure-cases} illustrates two representative tracking failures caused by such severe local adhesion.

In Case 1 (Fig.~\ref{fig:failure-cases}a), two adjacent catheters exhibit prolonged contact and similar local orientations. Within the adherent region, the candidate search neighborhoods of neighboring trackers substantially overlap, and the available foreground voxels provide insufficient information for reliable instance separation. Consequently, one tracker is gradually attracted toward the neighboring catheter and deviates from its expected trajectory.

In Case 2 (Fig.~\ref{fig:failure-cases}b), a tightly clustered catheter configuration causes one reconstructed trajectory to terminate before reaching its true endpoint. During tracking through the adherent region, the corresponding tracker fails to obtain sufficient validated candidate voxels for several consecutive iterations, possibly because the relevant foreground voxels are assigned to neighboring trackers or rejected by the directional and intensity continuity constraints. This results in premature trajectory termination.

\section{Discussion}\label{sec:discussion}

In this study, we developed a data-efficient and structure-aware framework for automatic multi-catheter digitization in CT-guided interstitial brachytherapy. By formulating the task as a physics-guided few-shot domain adaptation problem coupled with geometry-constrained tracking, the proposed method bridges the synthetic-to-real domain gap with only a 10-fraction annotation budget, enabling robust trajectory recovery in dense implant configurations.

A core methodological contribution of this study is the physics-driven source domain generation to overcome the intrinsic data scarcity in clinical settings. Rather than relying on heuristic data augmentations, we construct the source domain by integrating anatomically plausible template propagation with slice-wise metal artifact simulation. This generates highly realistic localized metallic signatures and artifact patterns, providing a robust pretext foundation for subsequent domain adaptation. Relying solely on synthetic supervision, IRNet achieved strong voxel-wise overlap but still exhibited occasional geometric outliers. With 10-shot fine-tuning, the HD95 was substantially reduced to 0.853 $\pm$ 0.362 mm, approaching the fully supervised clinical benchmark. This demonstrates that a limited number of real annotations is particularly valuable for correcting geometric outliers and improving downstream digitization reliability.

False-positive predictions scattered outside the implant region might have a minor impact on the Dice coefficient, but they severely disrupt line extraction, instance separation, and trajectory reconstruction. By incorporating implant region priors via auxiliary heatmap estimation and decoder modulation, IRNet effectively suppressed anatomically implausible background responses, such as distant cortical bone interference, and generated cleaner, more compact catheter masks. Therefore, for this application, the voxel-wise overlap and geometric reliability of the semantic mask should be considered comprehensively, rather than relying solely on the Dice metric.

While data-efficient segmentation accurately isolates the implant region, resolving these semantics into distinct instance-level trajectories represents the second major bottleneck. The performance improvements of the proposed reconstruction method are fundamentally linked to its precise grasp of the structural specificity of interstitial catheter reconstruction. Existing post-segmentation reconstruction strategies often rely on local continuity assumptions\cite{zhangMultineedleLocalizationAttention2020}, slice-by-slice tracking\cite{moradiFullyAutomaticReconstruction2025}, or region-of-interest digitization based on semantic masks. These strategies perform well when catheters are sparse and clearly separated, but they become less reliable when catheters touch, overlap, or intersect. Under these conditions, semantic segmentation fails to maintain instance identity, making the assignment of foreground voxels to individual catheters highly ambiguous. The proposed method addresses this challenge by combining global initialization with local physics-constrained tracking. The external catheter segments, which are less susceptible to tissue deformation, provide reliable directional information through a 3D Hough transform-based initialization. The subsequent synchronous inward tracking further incorporates geometric consistency and CT intensity continuity validation, mitigating erroneous switching between adjacent catheters. This global-to-local strategy accounts for the lower shaft and tip digitization errors and higher detection metrics observed in our method.

Experiments on the catheter-adhesion subgroup further corroborated the robustness of this framework. Adherent configurations are highly challenging because multiple catheters may exhibit extreme proximity or overlap across multiple consecutive axial slices, causing local or slice-by-slice separation mechanisms to fail completely. Although all methods experienced some performance degradation in this subgroup, our method maintained consistently superior accuracy and detection performance compared to the baselines. This indicates that the global direction prior and constrained inward tracking possess distinct advantages in handling strong local ambiguities. From a clinical perspective, this is crucial, as complex catheter interactions are prevalent in pelvic ISBT, and these complex cases represent the most time-consuming and error-prone scenarios for manual reconstruction.

This study has several limitations that warrant further investigation. First, while the framework demonstrates high geometric accuracy, its direct dosimetric impact remains unquantified. Given the steep dose gradients in HDR brachytherapy, future prospective studies must evaluate how these sub-millimeter residual errors affect clinical DVH parameters, such as target D90 and organ-at-risk D2cc. Second, the current cascaded paradigm inevitably propagates upstream false-negative omissions into the tracking module. Future algorithmic advancements will explore an end-to-end joint optimization framework, potentially integrating differentiable tracking mechanisms directly into the network architecture to allow bidirectional gradient flow. Finally, while the current simulation suffices for pretext training, it lacks complex biomechanical tissue deformations. Future efforts will incorporate comprehensive domain randomization strategies and multi-center validation to further enhance zero-shot or few-shot generalization across diverse catheter materials and implant templates.

\section*{Statements and Declarations}

\subsection*{Funding}
This work was supported by the Beijing Natural Science Foundation (No. L252005), the Key Specialty program of Natural Science Foundation of Beijing Municipality (Z20008 to P. Jiang).

\subsection*{Competing interests}
The authors have no conflicts of interest to declare.

\subsection*{Ethics approval}
The study was approved by the Ethics Committee of Peking University Third Hospital (approval No. IRB00006761-M2019210).

\subsection*{Data availability}
The data that support the findings of this study are available on request from the corresponding author. The patient data are not publicly available due to privacy or ethical restrictions. The code developed in this study is available from the corresponding author upon reasonable request.

\bibliographystyle{unsrt}
\bibliography{reference}

\end{document}